\documentclass[twocolumn]{aastex701}

\usepackage{subcaption}
\usepackage[italicdiff]{physics}
\usepackage[T1]{fontenc}

\def\msun{{M_\odot}}
\def\rsun{{R_\odot}}
\def\mearth{{M_\oplus}}
\def\rearth{{R_\oplus}}
\def\teff{{T_\text{eff}}}

\def\bigg{{\mathcal{G}}}

\newcommand\unif[1]{{\mathcal{U}[#1]}}
\newcommand\normal[1]{{\mathcal{N}(#1)}}
\newcommand\codestyle[1]{\textsc{#1}}

\def\vplanet{\codestyle{vplanet}}
\def\exofast{\codestyle{exofastv2}}
\def\batman{\codestyle{batman}}
\def\triceratops{\codestyle{triceratops}}
\def\astroimagej{\codestyle{astroimagej}}
\def\tapir{\codestyle{tapir}}
\def\banzai{\codestyle{banzai}}
\def\gaia{\textit{Gaia}}

\newcommand{\wiscaffil}{\affiliation{Department of Astronomy, University of Wisconsin--Madison, Madison, WI 53706, USA}}
\newcommand{\cfaaffil}{
\affiliation{Center for Astrophysics | Harvard \& Smithsonian, Cambridge, MA 02138, USA}
}

\newcommand{\revone}[1]{#1}

\begin{document}

\title{Discovery and Characterization of the TOI-4468 Planetary System: \\ A Transiting Hot Jupiter With a Lone Nearby Outer Companion}
\shorttitle{A Hot Jupiter With a Nearby Outer Companion}
\shortauthors{Livesey et al.}



\correspondingauthor{Joseph R. Livesey} 
\email{jrlivesey@wisc.edu}

\author[0000-0003-3888-3753]{Joseph R. Livesey}
\wiscaffil
\email{jrlivesey@wisc.edu}

\author[0000-0001-5084-4269]{Benjamin J. Hord} 
    \altaffiliation{NASA Postdoctoral Program Fellow}
    \affiliation{NASA Goddard Space Flight Center, Greenbelt, MD 20771, USA}
\email{benjamin.j.hord@nasa.gov}

\author[0000-0002-7733-4522]{Juliette Becker} 
\wiscaffil
\email{juliette.becker@wisc.edu}


\author[0000-0001-7246-5438]{Andrew Vanderburg} 
    \cfaaffil
\email{avanderburg@cfa.harvard.edu}

\author[0000-0001-8812-0565]{Joseph E. Rodriguez} 
    \affiliation{Center for Data Intensive and Time Domain Astronomy, Department of Physics and Astronomy, Michigan State University, East Lansing, MI 48824, USA}
\email{jrod@msu.edu}

\author[0009-0007-0740-0954]{Elise Koo}
    \affiliation{Anton Pannekoek Institute for Astronomy, University of Amsterdam, Amsterdam XH 1098, The Netherlands}
    \affiliation{ASTRON, Netherlands Institute for Radio Astronomy, Dwingeloo 7991 PD, The Netherlands}
\email{e.j.m.koo@uva.nl}

\author[0000-0003-0918-7484]{Chelsea X. Huang} 
    \affiliation{Centre for Astrophysics, University of Southern Queensland, Toowoomba, QLD 4350, Australia}
\email{chelsea.huang@usq.edu.au}

\author[0000-0001-7409-5688]{Guðmundur Stefánsson} 
    \affiliation{Astrophysics \& Space Institute, Schmidt Sciences, New York, NY 10011, USA}
    \affiliation{Anton Pannekoek Institute for Astronomy, University of Amsterdam, Amsterdam XH 1098, The Netherlands}
\email{gstefansson@schmidtsciences.org}

\author[0000-0001-6588-9574]{Karen A.\ Collins}
    \cfaaffil
\email{karen.collins@cfa.harvard.edu}


\author[0000-0003-0647-6133]{Ivan A. Strakhov} 
\affil{Sternberg Astronomical Institute, Lomonosov Moscow State University, Moscow 119992, Russia}
\email{strakhov.ia15@physics.msu.ru}

\author[0009-0009-6918-3259]{Zijun He} 
\affiliation{Madison West High School, Madison, WI 53726, USA}
\email{zijunhe66@gmail.com}

\author[0009-0002-2757-4138]{Maxwell A. Kroft} 
\wiscaffil
\email{mkroft@wisc.edu}


\author[0000-0003-2822-616X]{Robert Aloisi} 
    \wiscaffil
\email{rjaloisi@wisc.edu}

\author[0000-0003-1464-9276]{Khalid Barkaoui}
\affiliation{Instituto de Astrof\'isica de Canarias (IAC), La Laguna 38200, Tenerife, Spain}
\affiliation{Astrobiology Research Unit, Universit\'e de Li\`ege, Li\`ege 4000, Belgium}
\affiliation{Department of Earth, Atmospheric and Planetary Science, Massachusetts Institute of Technology, Cambridge, MA 02139, USA}
\email{khalid.barkaoui@uliege.be}

\author{Fabian Rodriguez Frustaglia}
\affiliation{Frustaglia Private Observatory, Spain}
\email{canopo100@hotmail.com}

\author[0000-0003-4287-004X]{Alyssa Jankowski} 
\wiscaffil
\email{alyssajanko5@gmail.com}

\author[]{Kinga Kasprzyk} 
\affiliation{Silesian University of Technology Observatory, Gliwice, Poland}
\email{kk306347@student.polsl.pl}

\author[0000-0002-0076-6239]{Judith Korth} 
\affiliation{Lund Observatory, Division of Astrophysics, Department of Physics, Lund University, Lund 22100, Sweden}
\email{judith.korth@fysik.lu.se}

\author[0009-0009-2277-799X]{Coleman Nelson} 
    \wiscaffil
\email{crnelson2250@gmail.com}

\author[0000-0001-5519-1391]{Hannu Parviainen} 
\affiliation{Departamento de Astrofísica, Universidad de La Laguna (ULL), La Laguna, Tenerife, Spain}
\affiliation{Instituto de Astrof\'isica de Canarias (IAC), La Laguna 38200, Tenerife, Spain}
\email{hpparvi@gmail.com}

\author[0000-0003-3184-5228]{Adam Popowicz}
\affiliation{Silesian University of Technology Observatory, Gliwice, Poland}
\email{apopowicz@polsl.pl}

\author[0000-0002-2190-3319]{Manfred Raetz}
\affil{Privat Observatory Herges-Hallenberg, Steinbach-Hallenberg, Germany}
\email{mraetz.herges@stiller-berg.de}

\author[0000-0001-8227-1020]{Richard P. Schwarz}
    \cfaaffil
\email{rpschwarz@comcast.net}

\author[0009-0001-2913-792X]{Eva Stafne} 
    \wiscaffil
\email{stafne@wisc.edu}

\author[0000-0002-3481-9052]{Keivan G. Stassun} 
    \affiliation{Vanderbilt University, Department of Physics \& Astronomy, Nashville, TN 37235, USA}
\email{keivan.stassun@vanderbilt.edu}

\author[0000-0002-9539-4203]{Thomas G. Beatty} 
\wiscaffil
    \email{tgbeatty@wisc.edu}

\author{Paul Foster} 
\affiliation{California State University, Los Angeles, Los Angeles, CA 90032, USA}
\email{paulfoster0923@gmail.com}

\author[0009-0001-6894-2657]{Thomas MacLean} 
\affiliation{California Institute of Technology, Pasadena, CA 91125, USA}
\affiliation{Department of Aeronautics and Astronautics, Stanford University, Stanford, CA 94305, USA}
\email{tmaclean@stanford.edu}

\author[0009-0009-0948-8819]{Devansh Mathur} 
    \wiscaffil
\email{dmathur2@wisc.edu}

\author[0000-0001-5401-8079]{Lizhou Sha} 
\wiscaffil
\affil{Department of Astrophysical Sciences, Princeton University, Princeton, NJ 08544, USA}
\email{lizhou.sha@princeton.edu}

\author[0000-0001-7493-7419]{Melinda Soares-Furtado} 
\wiscaffil
\email{mmsoares@wisc.edu}

\begin{abstract}


We report the discovery of two planets, a hot Jupiter and a nearby outer sub-Neptune, orbiting the star TOI-4468. This system is unique among the current exoplanet census in that it features a close outer companion to a hot Jupiter without an accompanying inner companion. By jointly fitting radial velocity measurements taken with the NEID spectrograph and transit photometry from TESS and several ground-based observatories, we constrain the orbital periods, masses, and radii of these two planets. We confirm the planetary nature of the hot Jupiter TOI-4468 b ($R = 1.01 R_J$, $m = 0.54 M_J$, $P = 2.77$ days). We also validate the outer planet TOI-4468 c ($R = 0.28 R_J$, $P = 7.01$ days) statistically, incorporating constraints from ground-based observations. We also identify, but cannot confirm, an additional radial velocity signal which may be due to an outer giant in this system \revone{with an orbital period of 624 days.} From the observed geometry of this system, we argue that it must never have encountered an early secular resonance that is thought to excite the mutual inclination of other hot Jupiter/outer companion systems. We discuss the possibility of an undetected inner companion, as well as potential implications for hot Jupiter formation.

\end{abstract}


\section{Introduction}
\label{sec:intro}
Constraining the formation mechanisms that produce hot Jupiters (HJs)---which, due to their ease of detection, are among the most abundant classes of exoplanets in the present observational sample despite being intrinsically rare \citep{Beleznay2022}---has been an ongoing effort in exoplanet astronomy since the discovery of the first specimen, 51 Pegasi b \citep{MayorQueloz1995}. Among the plausible formation channels are (i) \textit{in situ} formation, (ii) disk-driven migration, and (iii) high-eccentricity migration, driven either by a stellar flyby or planet--planet scattering in the outer system \citep{DawsonJohnson2018}. Channel (i) requires a large surface density of refractory material at the inner edge of the initial protoplanetary disk, which can \revone{possibly} be supplied in the presence of multiple short-period super-Earths \citep{Boley2016, Batygin2016}. A principal line of evidence favoring channel (iii) is that HJs tend to have outer giant companions, and these companions exhibit the same distribution of orbital periods as those in warm and cold Jupiter systems \citep{ZinkHoward2023}. Thus, HJs are connected to their outer systems, and do not typically co-migrate with their companions, as expected for disk-driven migration.

Channel (iii) likely precludes the existence of nearby exterior companions, as these planets would be driven unstable by dynamical interactions with the migrating HJ \citep{Mustill2015}, and also precludes the existence of inner companions as even ultra-short-period companions that might appear dynamically detached will not survive the migration process \citep{Becker2026}.
Several HJs have previously been discovered to have small inner companions. Almost all of these systems comprise solely the HJ and the inner companion: e.g.,
TOI-1130 \citep{Huang2020},
TOI-2000 \citep{Sha2023},
TOI-1408 \citep{Korth2024},
\revone{TOI-5143 \citep{Quinn2026,Radzom2025},}
TOI-2494 \revone{\citep{Quinn2026}}, 
Kepler-730 \citep{Cañas2019},
and WASP-132 \citep{Hord2022}.
WASP-47 is the only HJ system containing a nearby outer companion in addition to the inner \citep{Becker2015}. These HJs could not have formed through channel (iii), and thus either formed \textit{in situ} or migrated inward due to disk torques.
\begin{figure}
    \centering
    \includegraphics[width=\linewidth]{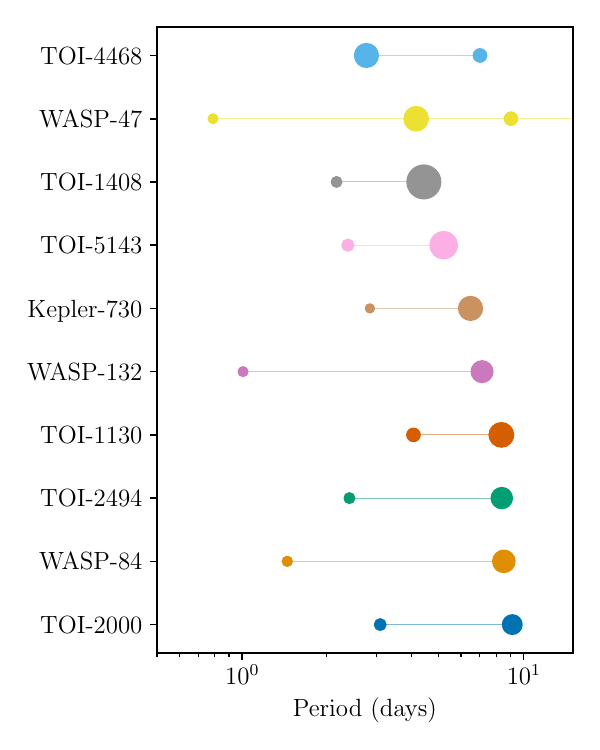}
    \caption{Known HJ systems containing nearby companions, with planetary radii to scale. WASP-47 b bears both an outer companion and \revone{an inner ultra-short period planet,} along with the cold giant WASP-47 c at 590 days. TOI-4468 is the first known system containing only a HJ and a nearby outer companion. The other eight systems exhibit only nearby inner companions.}
    \label{fig:companions}
\end{figure}

A recent analysis by \citet{Sha2026} determined a mean occurrence rate of 7.6\% for nearby companions to HJs. The census of HJs with nearby companions is presented in Figure \ref{fig:companions}.
The lack of observed nearby outer companions to HJs has been attributed to secular interactions between the two planets and the host star. In particular, as a low-mass star (below the Kraft break) loses angular momentum to its wind, it becomes less oblate as a consequence. The quadrupole moment of its gravitational field---parameterized by $J_2$, the leading-order zonal harmonic---decreases as a result. Consequently, the quadrupolar contribution to the planets' secular frequencies weakens, and when a critical value of $J_2$ is reached, a resonance between the secular frequencies and the instantaneous nodal regression rates of the planets is swept, amplifying their mutual inclination \citep{SpaldingBatygin2016, SpaldingBatygin2017, MacLeanBecker2025}. The expectation that this resonance---which acts only on two-planet systems in which the inner planet harbors significantly more angular momentum than the outer---affects the majority of HJ/outer companion systems explains the lack of such architectures in the observational census. Because the mutual inclination has been amplified in such a system, we will not see the companion to transit if we see the HJ.

In this work, we present the discovery and dynamical characterization of the TOI-4468 \revone{(TIC 441763252)} system. \revone{This system centers on a $\teff = 5200$ K K dwarf and} is the second observed, after WASP-47, that contains a HJ and a nearby outer companion, and is the first observed with \textit{only} an outer companion. This novel architecture presents the opportunity to further investigate the formation mechanism of HJs with nearby companions. In Section \ref{sec:data}, we explain the data used in our characterization of the system, including radial velocity measurements and transit photometry from multiple sources. We describe our analysis of the fitted data in Section, the confirmation of the HJ, and statistical validation of the companion in Section \ref{sec:analysis}. We discuss these results and consider the possibility of an additional inner companion in Section \ref{sec:discussion}, and conclude in Section \ref{sec:conclusions}.

\section{Data} \label{sec:data}

\subsection{TESS photometry}
TOI-4468 was observed by the Transiting Exoplanet Survey Satellite \revone{(TESS)}. TESS observes stars in periods of time known as sectors, each lasting about 28 days where the spacecraft maintains stable pointing at one part of the sky. Some regions of the sky, known as the ``continuous viewing zone'' are observed nearly continuously even as TESS's pointing moves from sector to sector. TOI-4468 falls near the continuous viewing zone, and as a result has been observed frequently over the course of the TESS mission. TOI-4468 has been observed by TESS for a total of 39 sectors as of November 2025. For the first year of observations (the second year of the TESS mission, sectors 14--26), TOI-4468 was not pre-selected for fast cadence observations, and instead was observed in the TESS full frame images, which were at that time coadded to 30-minute exposures. It was identified as a planet candidate host star by \citet{Olmschenk2021AJ} and was then selected for faster (2-minute) cadence observations in all sectors thereafter. After a considerable number of two-minute light curves were collected, a second planet candidate was identified and released by the TESS mission.

We reduced the data from the full frame images following \citet{Vanderburg2019ApJL}; in short, we extracted photometry from 20 different photometric apertures, corrected for systematics by decorrelating moments of the spacecraft quaternion time series and the background flux outside the aperture, and corrected for blending from outside the aperture. We used the standard TESS Science Processing and Operations Center \citep[SPOC;][]{tess-spoc} light curves for the two-minute cadence observations. We flattened both the 30-minute cadence full frame and two-minute cadence light curves following \citet{Pepper2020AJ} and used these light curves in our global modeling. \revone{Our full frame images were taken in sectors 14--26, and 2-minute cadence data from sectors 41, 48--60, 73, and 75--80.}

\subsection{Ground-based photometry\label{sec:groundtransits}}

The TESS pixel scale is $\sim 21''$ pixel$^{-1}$ and photometric apertures typically extend out to roughly 1 arcminute, generally causing multiple stars to blend in the TESS photometric aperture. To determine the true source of the TESS detections, we acquired ground-based time-series follow-up photometry of the field around TOI-4468 as part of the TESS Follow-up Observing Program Sub Group 1 \citep[TFOP SG1;][]{collins:2019}.\footnote{\href{https://tess.mit.edu/followup}{https://tess.mit.edu/followup}} We used the TESS Transit Finder, which is a customized version of the \tapir~software package \citep{Jensen:2013}, to schedule our transit observations. 

We used \astroimagej~\citep{Collins:2017} to extract differential photometric data from five TOI-4468 b image sequences and one TOI-4468 c image sequence, as detailed below. The light curve data were extracted using circular photometric apertures having radii ranging from $5\farcs2$ to $10\farcs1$. The target star apertures excluded all flux from all known Gaia DR3 catalog neighbors. Transit-like signals were detected in the target star apertures in all six light curves, confirming that both TESS detected transit-like events are indeed occurring at TOI-4468. The light curve data are available on the ExoFOP-TESS website.\footnote{\href{https://exofop.ipac.caltech.edu/tess/target.php?id=441763252}{https://exofop.ipac.caltech.edu/tess/target.php?id=441763252}} The light curves are included in the global model described in Section \ref{sec:globalmodel} and are shown in Figure \ref{fig:tfop-transits}.

\subsubsection{Herges-Hallenberg 0.28\,m} 
We observed a transit egress window of TOI-4468 b with no filter on UTC 2022 March 18 from the Privat Observatory Herges-Hallenberg 0.28\,m telescope near Steinbach-Hallenberg, Germany. The telescope is equipped with a Moravian Instrument G2-1600 detector with an image scale of $1\farcs$02 pixel$^{-1}$, resulting in a $27\arcmin\times41\arcmin$ field of view. Images were calibrated using \astroimagej.

\subsubsection{Frustaglia Observatory 0.3\,m} 
We observed three full transit windows of TOI-4468 b on UTC 2023 January 11, 2023 April 07, and 2025 December 17 from the Frustaglia Private Observatory 0.3\,m telescope located near Fregenal de la Sierra, Spain (RFAC). The telescope is equipped with a $6248\times4176$ QHY 268M camera having an image scale of $0\farcs32$ per pixel, resulting in a $33\arcmin\times22\arcmin$ field of view. The images were calibrated using \astroimagej.

\subsubsection{LCOGT 1.0\,m} 
We observed a full transit window of TOI-4468 c on UTC 2024 August 16 and a full transit window of TOI-4468 b on UTC 2025 April 25, both in Sloan $i'$ band, from the Las Cumbres Observatory Global Telescope \citep[LCOGT;][]{Brown:2013} 1.0\,m network node at Teide Observatory on the island of Tenerife. The 1\,m telescopes are equipped with $4096\times4096$ SINISTRO cameras having image scale $0\farcs389$ $\rm{pixel}^{-1}$, resulting in $26\arcmin\times26\arcmin$ field of views. The images were calibrated by the standard LCOGT \banzai~pipeline \citep{McCully:2018}.


\subsubsection{Silesian University of Technology Observatory}
We observed a full transit on UT 2022 March 09 from the Silesian University of Technology Observatory (SUTO) 0.3\,m telescope located near Otivar, Spain. The telescope is equipped with a $4656\times3520$ pixel ASI ZWO 1600MM camera. The image scale is 0$\farcs$685 pixel$^{-1}$, resulting in a $53\arcmin\times42\arcmin$ field of view. The photometric data were extracted with \astroimagej~using a circular $7\farcs2$ aperture.

\subsection{Radial velocities}
\begin{deluxetable}{lcc}
\tablecaption{Radial velocities measured with NEID.
\label{tab:rvs}
}
\tablehead{
   \colhead{Epoch (BJD$_\text{TDB}$)} & \colhead{Radial velocity (m/s)} & \colhead{Error (m/s)}
}
\startdata
2460194.68192 & $6.2$ & 7.4 \\
2460236.66439 & $-82.9$ & 9.7 \\
2460239.59063 & $-101.96$ & 8.7 \\
2460327.99483 & $-58.0$ & 6.3 \\
2460329.00208 & $-107$ & 11 \\
2460341.94347 & $-69$ & 10 \\
2460411.96062 & $-99.8$ & 8.1 \\
2460447.70677 & $-98$ & 10 \\
2460454.76181 & $84.7$ & 7.5 \\
2460472.83070 & $-87.5$ & 6.2 \\
2460595.62048 & $58.7$ & 7.1 \\
2460753.93099 & $63.4$ & 5.8 \\
2460755.92351 & $-46.6$ & 6.2 \\
\enddata
\end{deluxetable}
Our RV measurements were taken with NEID, an ultra-precise stabilized \citep{Stefansson2016, Robertson2019}, fiber-fed \citep{Kanodia2023} spectrograph on the WIYN 3.5 m Telescope. \revone{NEID has continuous broadband wavelength coverage between 380--930 nm. Our observations used the high-efficiency mode, with spectral resolving power $R = 60,000$.}
We have 13 total data points each with an exposure time of roughly 3300 seconds, which were taken between Summer 2023 and Spring 2025. The individual measurements are given in Table \ref{tab:rvs}.

The radial velocities were extracted with the spectral-matching method, using a code inspired by SpEctrum Radial Velocity AnaLyzer \citep[SERVAL;][]{Zechmeister2018}, but adapted for NEID \citep{Stefannsson2022}. \revone{As this pipeline is optimized for faint targets it is useful for TOI-4468, which has an apparent $V$-band magnitude of 14.2.} These extracted SERVAL RVs agree well with the Cross-Correlation Function (CCF) RVs extracted through the standard NEID pipeline, while reducing the median error from 91.49 m/s to 8.11 m/s. 
\revone{Our observations have signal-to-noise ratios ranging between 10 and 18, and were taken at airmasses between 1.32 and 1.95.}

\subsection{Stellar spectral energy distribution}
We adopt the SED values reported in ExoFOP for TOI-4468 as our priors, as listed in Table \ref{tab:sed}. \revone{These values were derived in the TIC catalog following the methods described in \citet{Stassun2018}.} The output SED from our \exofast~global fit of the system (Section \ref{sec:globalmodel}) is shown in Figure \ref{fig:sed}.
\begin{deluxetable}{rcc}
\tablecaption{Spectral energy distribution of TOI-4468 in magnitudes.
\label{tab:sed}
}
\tablehead{
   \colhead{Band} & \colhead{Magnitude} & \colhead{Error (mag)}
}
\startdata
    $B$ & 14.86 & 0.12 \\
    $V$ & 14.08 & 0.20 \\
    Gaia $G$ & 13.8677 & 0.0004 \\
    $g'$ & 15.1888 & 0.0039 \\
    $r'$ & 13.9853 & 0.0026 \\
    $i'$ & 14.2862 & 0.0053 \\
    $z'$ & 13.7478 & 0.0047 \\
    2MASS $J$ & 12.453 & 0.020 \\
    2MASS $H$ & 11.977 & 0.019 \\
    2MASS $K$ & 11.890 & 0.024 \\
    WISE 3.4 $\mu$m & 11.873 & 0.022 \\
    WISE 4.6 $\mu$m & 11.936 & 0.021 \\
    WISE 12 $\mu$m & 11.97 & 0.17
\enddata
\end{deluxetable}


\subsection{Speckle imaging}
\begin{figure}
    \centering
    \begin{subfigure}{\linewidth}
        \centering
        \includegraphics[width=\linewidth]{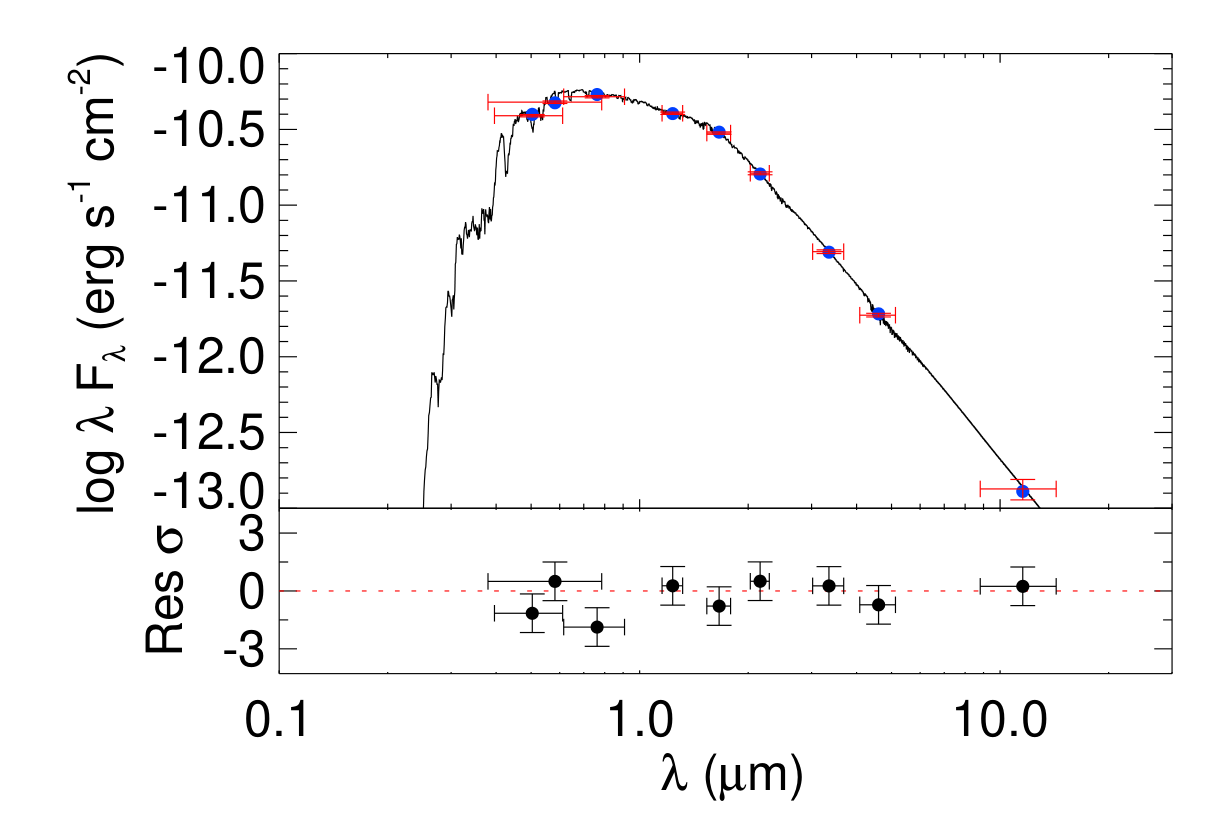}
        \caption{Spectral energy distribution of TOI-4468 in units of flux, as calculated in our global fit with \exofast.}
        \label{fig:sed}
    \end{subfigure}
    \vspace{-0.5cm}
    
    \begin{subfigure}{\linewidth}
        \centering
        \includegraphics[width=\linewidth]{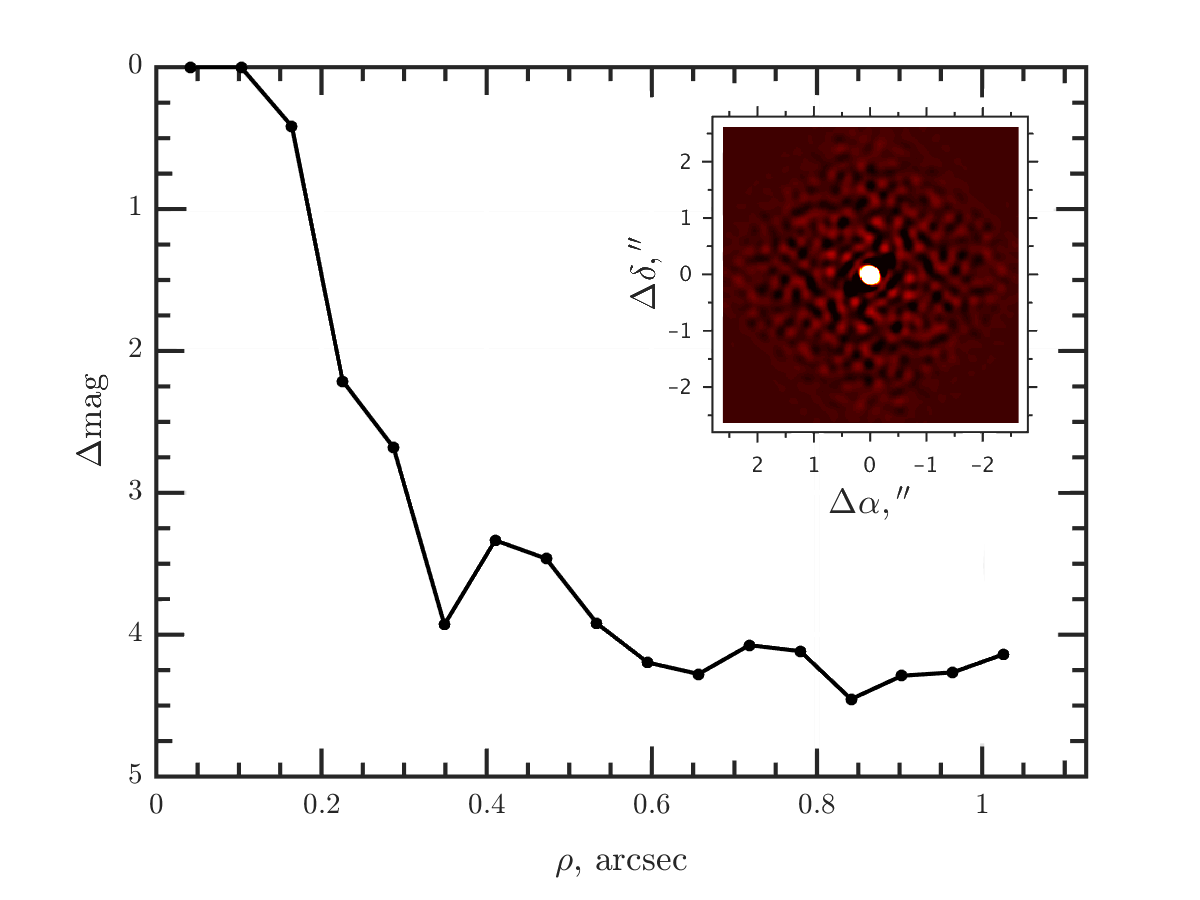}
        \caption{Detection limits for TOI-4468 obtained using speckle interferometry at SAI-2.5m telescope. The inset contains the autocorrelation function of the object. North is up, east is to the left.}
        \label{fig:speckle}
    \end{subfigure}
    \caption{(a) Spectral energy distribution of TOI-4468. (b) Speckle interferometry detection limits.}
    \label{fig:combined}
\end{figure}

We observed TOI-4468 on 28th February, 2024 using speckle interferometry with the Speckle Polarimeter (SPP), a facility instrument on the 2.5 m telescope at the Caucasian Observatory of the Sternberg Astronomical Institute (SAI) of Lomonosov Moscow State University \citep{Strakhov2023}. We used a fast low-noise CMOS Hamamatsu ORCA-quest as a detector and $\mathrm{I_c}$ filter. The pixel scale was 20.6 mas/pixel. The power spectrum was estimated from 2500 frames per star with 60 ms exposure time, and is shown in Figure \ref{fig:speckle}. The atmospheric dispersion compensator was employed. We did not detect a stellar companion, the detection limits are $3.1$ mag and $4.2$ mag at 0\farcs25 and 1\farcs0 from the star, respectively.


\section{Analysis} \label{sec:analysis}
\subsection{Global system characterization}
\label{sec:globalmodel}
We use the \exofast~code \citep{Eastman2019} to self-consistently model the star and both planets in the TOI-4468 system, constraining the planetary properties by jointly fitting transit data from TESS, RFAC, LCOGT, and SUTO and RVs measured with the NEID spectrograph. \revone{The TESS data comprise 311 transits of planet b in total, including 8 partial transits.} We do not fit RVs for planet c, as its signal is below the noise threshold for NEID ($3.85$ m/s for a mass of $10\mearth$). Aside from an uninformative uniform prior for the dust extinction, our priors are taken from \gaia. They are listed in Table \ref{tab:priors}.

\begin{deluxetable*}{llc}
\tablecaption{Priors used to initialize the MCMC fit.\label{tab:priors}}
\tablehead{
   \colhead{Parameter} & \colhead{Description} & \colhead{Prior}
}
\startdata
[Fe/H] & Metallicity & $\normal{0.0, 0.25}$ \\
$A_V$ & $V$-band extinction (mag) & $\unif{0, 0.75}$ \\
$\varpi$ & Parallax (mas) & $\normal{2.645, 0.0134}$ \\
\enddata
\tablecomments{All priors not listed in this table are set to the \exofast~defaults, \revone{which are uninformative uniform priors \citep[see][Table 3]{Eastman2019}.}}
\end{deluxetable*}


\revone{We verify the convergence of our \exofast~fit by running it up to a Gelman--Rubin statistic \citep{GelmanRubin1992} less than 1.01 \citep[see e.g.,][]{Eastman2019}.} We find mean planetary radii of $1.01 R_J$ and $0.28 R_J$ for planets b and c, respectively, along with a mean mass of $0.54 M_J$ for the former. The equilibrium temperature of planet b is roughly 1150 K. We find an eccentricity consistent with zero for planet b.

\begin{figure}
    \centering
    \includegraphics[width=\linewidth]{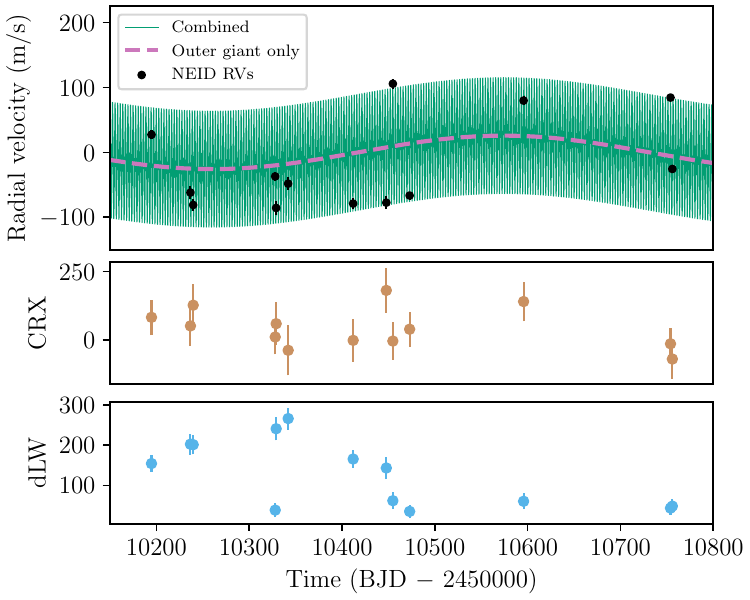}
    \caption{Stellar activity indicators for TOI-4468 measured by NEID: the chromatic index (CRX) and differential line width (dLW). There appears to be some evidence of variation in dLW on the same period as the RV signal attributed to TOI-4468.03. Without more data, however, we cannot verify whether the additional RV signal is due to stellar activity or to a third planet in the system.}
    \label{fig:activity-indicators}
\end{figure}

We identify a third signal present in the RVs which may be due to a giant planet on a longer-period orbit. Our measurements poorly constrain the semi-amplitude of this additional signal, and our baseline is shorter than its apparent period. We cannot conclusively rule out stellar activity as the source of this additional signal; stellar activity indicators are shown in Figure \ref{fig:activity-indicators}. We fit this signal with \exofast~and find that, if it is of planetary origin, it belongs to a companion planet with an orbital period of 624 days. \revone{We cannot well constrain its minimum mass, but find a median value of $0.98 M_J$. We discuss the possibility and implications of this third planet further in Section \ref{sec:discussion}.}

\begin{figure*}
    \centering
    \includegraphics[width=\linewidth]{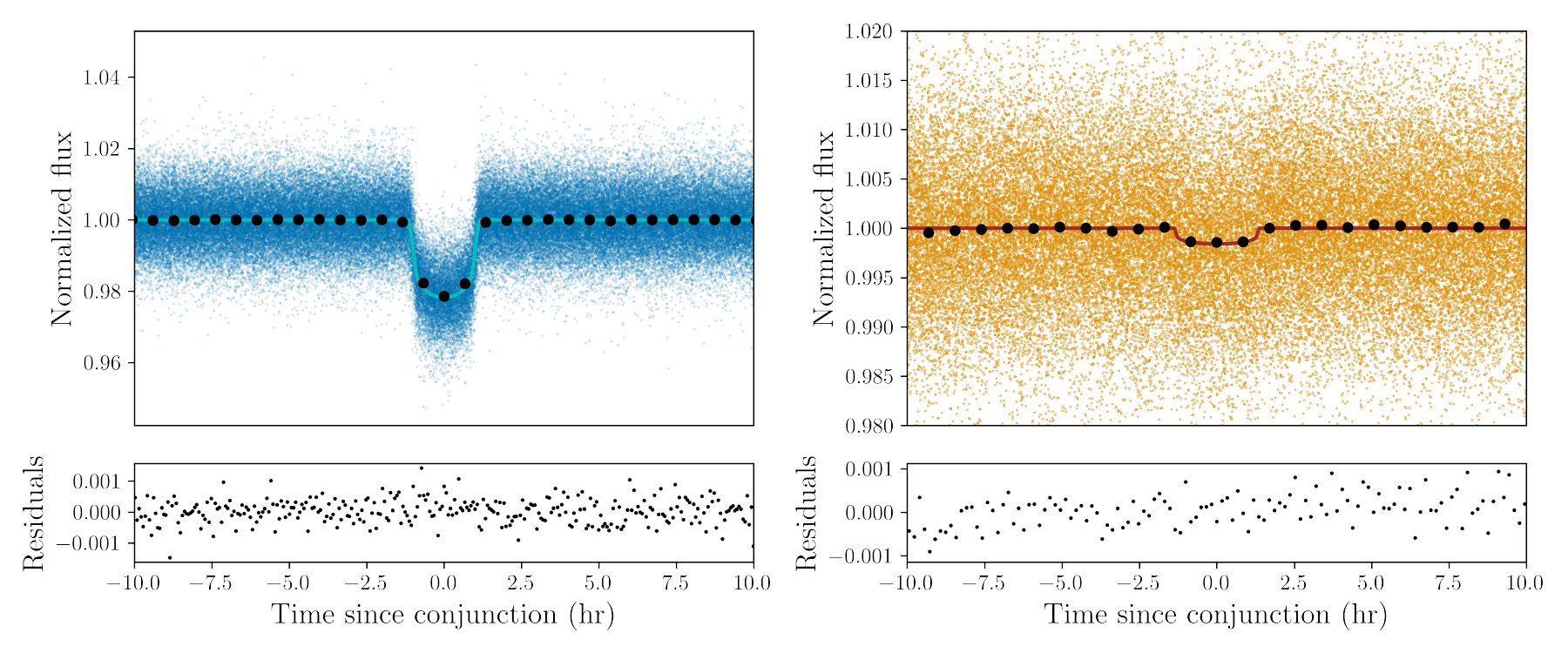}
    \caption{Transit model light curves for both planets, laid over the raw photometric measurements taken with TESS. The residuals are binned by a factor of 500. \textbf{Left:} TOI-4468 b; \textbf{Right:} TOI-4468 c.}
    \label{fig:transits}
\end{figure*}

\begin{figure}
    \centering
    \includegraphics[width=\linewidth]{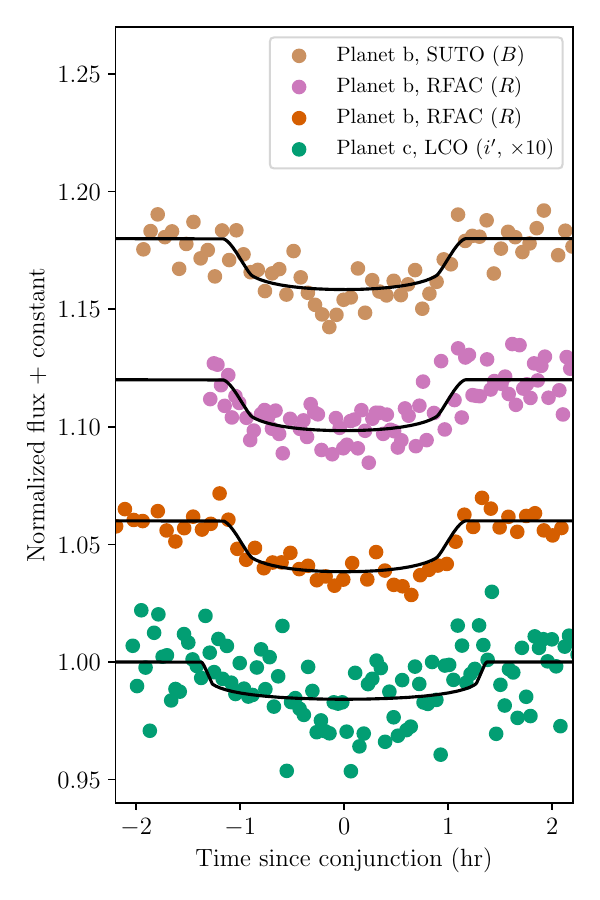}
    \caption{Transit data taken from ground-based observatories. The data from SUTO, and RFAC exhibit transits of planet b, while the transit from LCO is of planet c (the difference in normalized flux for this transit is magnified here by a factor of 10 for visibility). These data are included in our global fit with \exofast.}
    \label{fig:tfop-transits}
\end{figure}

\begin{figure*}
    \centering
    \includegraphics[width=\linewidth]{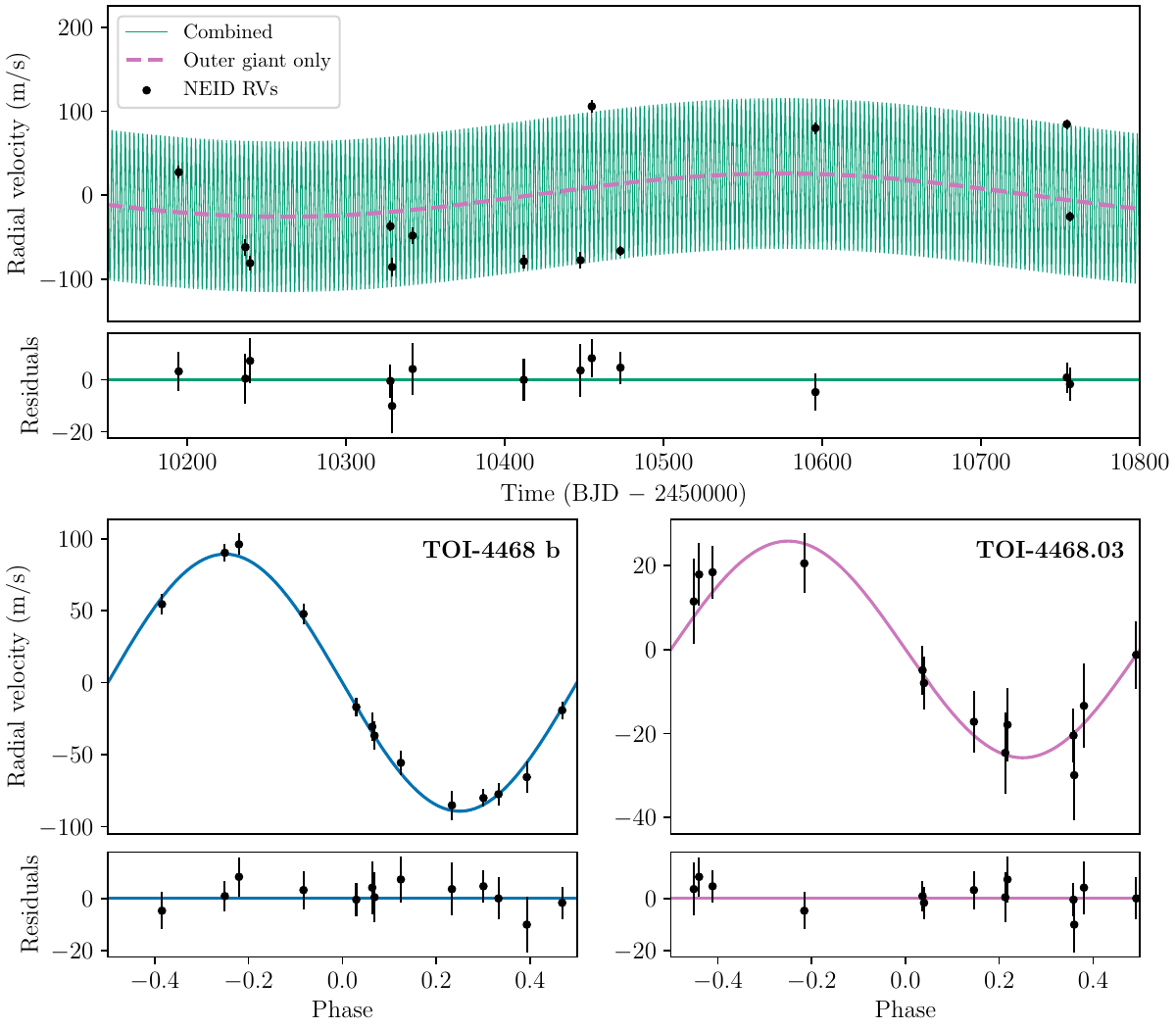}
    \caption{Best-fit radial velocity curves for the giants TOI-4468 b and the unconfirmed TOI-4468.03, both individually and combined. The best-fit eccentricity of planet b is $e_b = 0.030$.}
    \label{fig:rv}
\end{figure*}

The full list of posteriors is provided in Tables \ref{tab:exofast} and \ref{tab:exofast-photometry}. In Figure \ref{fig:transits} we show the raw transit data from TESS, and overlay the model light curves for planet b (left panels) and planet c (right panels) derived from our fit with \exofast, generated with the \batman~package \citep{Kreidberg2015}. In the transit plot for planet b (c) we have removed all data from during any transit of planet c (b). In Figure \ref{fig:tfop-transits} we plot three more transits of planet b and one of planet c, taken from ground-based observatories, that we include in our global fit. In Figure \ref{fig:rv} we plot the raw RVs with the model derived from our \exofast~fit shown.

\subsection{Statistical validation of TOI-4468 c}

While vetting and observational constraints can rule out many astrophysical false positive sources, these observational limits often cannot rule out the entire false positive parameter space. To determine the likelihood that the 7-day signal is from an astrophysical source in the false positive parameter space, we perform a statistical analysis using the publicly available software \triceratops~\citep{triceratops-code, triceratops-paper}, using the available photometry and imaging observations of TOI-4468 as constraints.

\triceratops~compares the photometric signal to a suite of astronomical false positive sources including eclipsing binaries, background binaries, background transiting planets, nearby eclipsing binaries, transiting planets around an unresolved bound companion star, and double periods of each. The software was specifically designed for use with TESS observations, accounting for the light curve extraction aperture, the nearby stars contained within, and all stars within $2\farcm5$ of the target when calculating the false positive probability of the signal (FPP) and likelihood that the signal is a false positive originating from a nearby star (the nearby FPP, or NFPP).

To perform the FPP analysis with \triceratops, we used the TESS SPOC PDC$\_$SAP two-minute cadence light curves for all available TESS sectors. Since \triceratops~accounts for the TESS extraction aperture in the FPP calculation, we used the PDC$\_$SAP apertures provided by the SPOC where applicable. We also incorporated the $I$-band contrast curve from the SPP instrument and the constraints on potential nearby stars as the source of the signal from multi-band photometric follow up.

The TESS light curves were phase-folded using the values from the global fitting found in Table \ref{tab:exofast}. We ran the FPP calculation with \triceratops~20 times and took the average FPP value. This results in a $\text{FPP} = 0.041 \pm 5.32 \times 10^{-4}$ for the 7-day signal. This is just short of the FPP $<$ 0.01 necessary to statistically validate the signal as a planet, however this does not take into account the fact that there is another planet in the system. Signals in systems with multiple transiting planet candidates are considerably more likely to be genuine transiting exoplanets than those in single candidate systems \citep{Latham2011, Lissauer2011, Guerrero2021}. We are therefore able to apply this ``multiplicity boost'' to the FPP of the 7-day signal. \citet{Guerrero2021} quantified the multiplicity boost of candidates discovered by TESS to be 20$\times$ for systems containing planets larger than $6 R_{\oplus}$. Applying this boost to our the FPP from \triceratops, we obtain $\text{FPP} = 2.05 \times 10^{-3}$, well below the $\text{FPP} = 0.01$ threshold necessary for statistical validation. Therefore, we consider TOI-4468 c to be a bona fide planet.

We note that we do not perform a separate statistical validation for the inner, 2.8-day signal since not only was a $>5\sigma$ mass measurement obtained for that planet, but also statistical validation for planets $R_{p} > 8 R_{\oplus}$ may not be valid due to the overlap in parameter space with brown dwarfs.

\subsection{Transit timing variations}
Systems containing hot Jupiters and nearby companions have been previously observed to have significant transit timing variations \citep[e.g.,][]{Huang2020} due to planet-planet interactions. These interactions are stronger for planet pairs near a mean motion commensurability. The period ratio between TOI-4468 c and b is roughly 2.53, slightly exterior to the 5:2 commensurability. To test whether this geometry produces observable planet–planet interactions, we perform a one-parameter fit using \batman~to each individual transit in the 2-minute cadence data, holding all system parameters fixed to the best-fit solution listed in Table \ref{tab:exofast} and allowing only the mid-transit time and continuum shape to vary. 
Transits lacking at least 2.0 transit durations of data centered on the transit time, as well as transits involving overlapping events from both planets within two durations, were excluded from the fit. This resulted in the exclusion of 11 events for TOI-4468 b and 11 events for TOI-4468 c.

Fits to a flat-line model yield reduced $\chi^2$ values of 2.1 and 1.8 for TOI-4468 b and c, respectively, suggesting either the presence of a small amount of excess timing scatter beyond the formal uncertainties or that our uncertainties are underestimated.
To check the possible TTV amplitudes that would be compatible with the data, we compute the weighted root-mean-square of the residuals of the O–C values.
This yields values of 3.4 minutes for TOI-4468 b and 24.8 minutes for TOI-4468 c, corresponding to 95\% upper limits on possible TTV amplitudes of 6.8 minutes and 48.6 minutes, respectively.
TTV amplitudes below these thresholds are consistent with the observed transit times, whereas signals exceeding these levels are ruled out by the data.

To examine possible long-term TTV trends in the 
TTV signal of TOI-4468 b, we performed an additional fit for the center times of transit computed for the data from each individual TESS sector, using \codestyle{juliet} \citep{Espinoza2019} to fit the period $P$, center time of transit $T_{0}$, $R_p/R_\star$, $a/R_\star$, $\sqrt{e} \cos(\omega)$, and $\sqrt{e} \sin(\omega)$. A single $T_{0}$ was computed for each sector of TESS data. The resultant computed $T_{0}$ values are consistent to $\sim1\sigma$, providing no additional evidence of sector-to-sector timing variations.

\section{Discussion} \label{sec:discussion}
\subsection{The possibility of an inner companion}
TOI-4468 b is to date the only known hot Jupiter accompanied by a small outer planet but no inner planet. The only other hot Jupiter with an identified outer companion, WASP-47 b, also harbors a small interior planet \citep{Becker2015}.
If this system had an inner companion and formed during the disk phase, it should (for most parameter combinations) reside close enough to the orbital plane of the HJ to transit as well \citep{MacLeanBecker2025}. 

The 2.8-day orbit of planet b sets a narrow limit inside which a potential interior companion could reside. Therefore the natural first question to address is whether it is possible to place a planet here. We adopt the classic stability criterion due to \citet{Gladman1993}: that $a - a_b \geq 2\sqrt{3} R_H$, where
\begin{align} \label{eq:chambers}
    R_H = \frac{a + a_b}{2} \left ( \frac{m + m_b}{3M_\star} \right )^{1/3}
\end{align}
is the mutual Hill radius of planet b and the hypothetical companion.

\begin{figure}
    \centering
    \includegraphics[width=\linewidth]{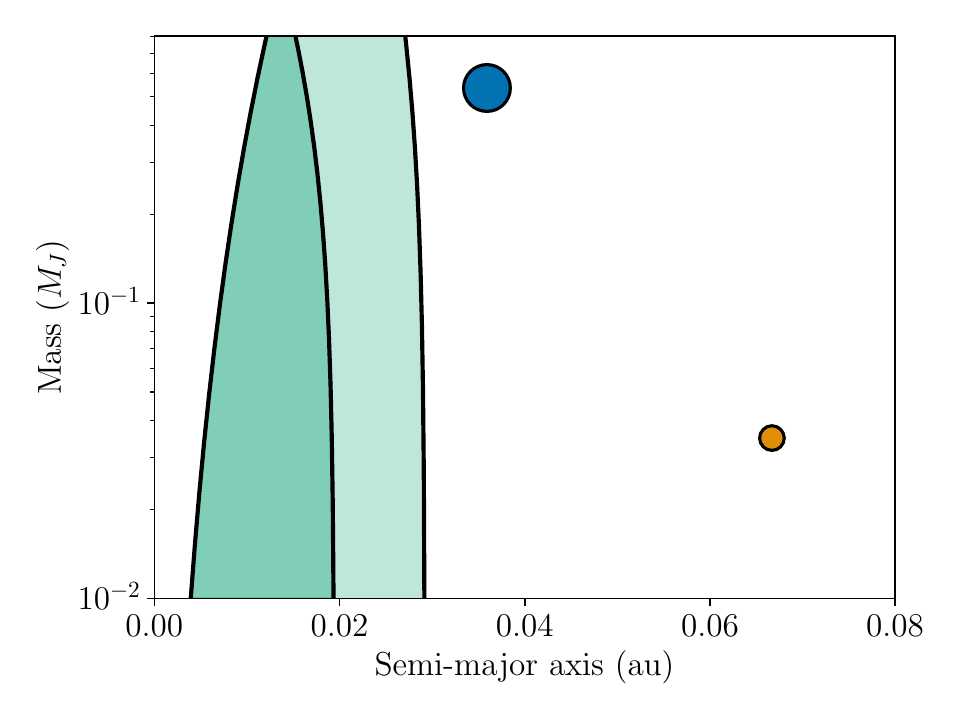}
    \caption{Stability limits for a possible inner companion to TOI-4468 b. The innermost boundary is the Roche limit. The dark green area is the conservatively stable regime (beyond $10 R_H$ from planet b). The light green area is the Hill stable regime (beyond $2\sqrt{3} R_H$ from planet b). The locations of the two confirmed planets in this parameter space are shown for reference.}
    \label{fig:inner-planet}
\end{figure}

In Figure \ref{fig:inner-planet}, we show the calculated stability limits for an inner planet. The overlain contours are the $2\sqrt{3} R_H$ criterion of \citet{Gladman1993} and the more stringent $10 R_H$ criterion derived from numerical simulations of two-planet systems \citep[e.g.,][]{Obertas2017}. The inner boundary on the allowed region is the Roche limit. The Roche radius is
\begin{align}
    R_R = \left ( \frac{2\rho_\star}{\rho} \right )^{1/3} R_\star,
\end{align}
where $\rho_\star = 2.48$ g cm$^{-3}$ based on our fit values for the stellar mass and radius. The planetary density is calculated assuming a spherical planet with radius related to its mass \revone{by the prescription of \citet{ChenKipping2017}:}
\begin{align}
    R = \begin{cases} \rearth (m / \mearth)^{0.28}, & m < 2\mearth \\ 2^{0.28} \rearth (m / 2\mearth)^{0.59}, & m \geq 2\mearth. \end{cases}
\end{align}
A large swath of parameter space interior to the hot Jupiter is amenable to hosting a small planet; we cannot in principle rule out the presence of such a companion.

Using a simplistic scaling argument, we can determine approximately the minimum radius at which a transiting planet would have avoided detection as a function of orbital period. \revone{The detection significance $S$} scales to a first-order approximation with $R^2 / \sqrt{P t_\text{dur}}$, where $t_\text{dur}$ is the duration of the transit \citep[e.g.,][]{Sha2026}. The transit duration scales with period as $t_\text{dur} \propto P^{1/3}$ by Kepler's third law. The threshold radius therefore follows
\begin{align}
    R \simeq \left ( \frac{P}{P_c} \right )^{1/3} \left ( \frac{S}{S_c} \right )^{1/2} R_c.
\end{align}
Using the values we report here for the radius and orbital period of planet c, as well as its detection significance ($S_c \sim 25.2\sigma$) we can compute the requisite radius for a $7\sigma$ detection at the orbital period of planet b. We find that this radius is roughly \revone{$1.2 \rearth$}, indicating that in the stable regime we have identified an undetected inner companion would have a size of at most the order of Earth's.

\subsection{The small mutual inclination}
A likely cause for the dearth of small outer companions to hot Jupiters is the crossing of a secular resonance early in such a system's lifetime \citep[e.g.,][]{MacLeanBecker2025}. This resonance arises when the ratio of two combinations of the secular frequencies is unity \citep{SpaldingBatygin2016}. Early on, the precession of both planes is primarily due to coupling with the stellar quadrupole, and therefore the star's $J_2$ controls whether the system is in this resonance. The resonance is swept when
\begin{align} \label{eq:J2}
    J_2 \simeq \frac{1}{6} \frac{m_2}{M_\star} \frac{(\Lambda_1 + \Lambda_2)^2}{(\Lambda_1 - \Lambda_2) \Lambda_2} \left ( \frac{a_1}{R_\star} \right )^2 \frac{\alpha^2 b^{(1)}_{3/2}(\alpha)}{1 - \alpha^{7/2}},
\end{align}
where $\alpha \equiv a_1/a_2$ and $\Lambda_i \equiv m_i \sqrt{\bigg M_\star a_i}$ is the circular angular momentum of the $i$-th planet. Here the function
\begin{align}
    b_{3/2}^{(1)}(\alpha) = \frac{2}{\pi} \int_0^\pi \frac{\cos(\psi) \: d\psi}{[1 + \alpha^2 - 2\alpha \cos(\psi)]^{3/2}}
\end{align}
is the usual Laplace coefficient \citep{MurrayDermott1999}. Note that this value changes as the star evolves; in particular, it increases as the star contracts during the pre-main sequence phase.

\begin{figure}
    \centering
    \includegraphics[width=\linewidth]{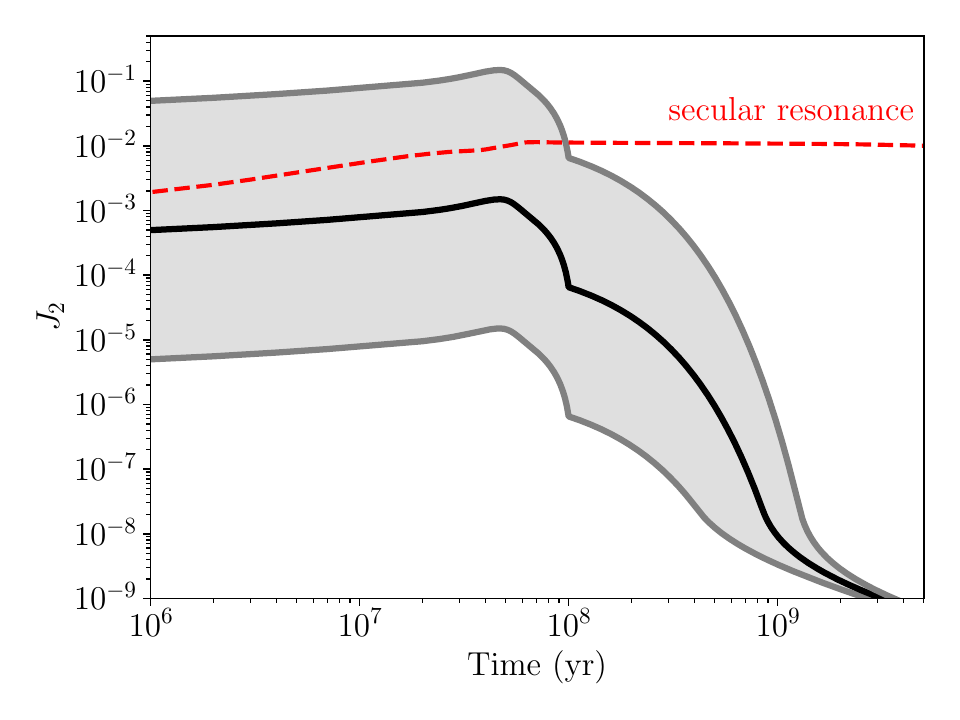}
    \caption{Estimated temporal evolution of TOI-4468's $J_2$ for a range of initial rotation periods, $0.1 < P_\text{rot} < 10$ days. The necessary $J_2$ to achieve a secular resonance between planets b and c is highlighted. The case with an initial rotation period of 1 day is traced by the black curve. $J_2$ may reach the resonant value for initial rotation periods significantly less than 1 day. It is therefore unlikely, but not impossible, that this resonance was ever crossed in the system's history. The kink at $10^8$ years is due to $k_2$ becoming constant as the star's interior becomes fully radiation-dominated. At ages $> 1$ Gyr, $k_2$ and $\Omega_b$ are roughly constant, so $J_2 \propto t^{-1}$ per the braking law of \citet{Skumanich1972}.}
    \label{fig:J2}
\end{figure}

Using the values we fit for the system in Section \ref{sec:analysis}, we find that the quadrupole moment necessary to trigger the resonance has varied over the course of the star's life from $J_2 \sim 10^{-3}$--$10^{-2}$. This range corresponds to an extremely oblate star, and is unlikely to have been achieved at any point during the system's history. The time dependence of the stellar $J_2$ is implicit in the relation
\begin{align}
    J_2 = \frac{1}{3} k_2 \left ( \frac{\Omega_\star}{\Omega_b} \right )^2,
\end{align}
where $k_2$ is the star's degree-2 Love number, $\Omega_\star$ is the angular velocity at the surface along its equator and $\Omega_b = \sqrt{\bigg M_\star/R_\star^3}$ is the breakup angular velocity. We use the \citet{Matt2015} model of magnetic braking to compute the spin evolution of the star early in its life. This model applies a time-dependent wind torque to a star whose instantaneous physical state is interpolated from the pre-computed evolutionary grids of \citet{Baraffe2015}. Using the stellar evolution module in the \vplanet~code \citep{Fleming2019, Barnes2020}, we compute $\Omega_\star(t)$ in this way up to an age of 5 Gyr. We also compute the evolution of the stellar radius to get $\Omega_b(t)$. Following \citet{Becker2020}, we adopt the following simple piecewise continuous evolution for the Love number as the stellar interior transitions from convection-dominated to radiation-dominated.
\begin{align}
    k_2(t) = \begin{cases}
        0.28, & t < 10^7 \text{ yr} \\
        a + b t, & 10^7 \leq t \leq 10^8 \text{ yr} \\
        0.014, & t > 10^8 \text{ yr}
    \end{cases}
\end{align}
In the above, $a = 0.309566$ and $b = -2.95556 \times 10^{-9}$ yr$^{-1}$. Having obtained the time dependence of $\Omega_\star$ and $k_2$, we calculate the evolution of TOI-4468's $J_2$ over its lifetime. The result is plotted in Figure \ref{fig:J2}. Clearly the stellar $J_2$ never reaches the required value to trigger this resonance.
Because it is unlikely that the system ever encountered the resonance that would evict planet c from the invariable plane, the fact that the mutual inclination between the two planets is small tells us nothing about the stellar obliquity. That this resonance was avoided, however, means that it remains a compelling explanation for the loneliness of HJs, in spite of TOI-4468's coplanarity.


We note also that in the limit where the outer planet is a test particle, even a modest stellar obliquity guarantees a large final mutual inclination between the HJ and its companion, due to adiabatic change in the phase space topography as the stellar quadrupole moment decreases \citep{Batygin2016}. The planetary mass ratio $m_c/m_b$ in the TOI-4468 system is large enough, however, that it is better treated by the previously detailed two-massive-planet model of \citet{SpaldingBatygin2016}.

\subsection{The third RV signal, and the possible effect of an outer giant planet}
There is a clear RV signal with a semi-amplitude of 25.8 m/s and a period of $\sim 624$ days present in our data (lower right panel, Figure \ref{fig:rv}). With the data available to us, we are not able to determine if this signal is due to stellar activity. Stellar activity indicators (Figure \ref{fig:activity-indicators}) show some degree of correlation with the $\sim 624$ signal, but the quantity of RV data available is insufficient to prove either that the signal is due to stellar activity. For this reason, we must await future RV data, which will allow for a determination of whether TOI-4468.03 is a bona fide planet or not. 
If this signal is evidence of an additional outer giant in the system, and not stellar activity, then this raises some new dynamical implications. 

The possible presence of a cold Jupiter in the system could contribute an additional quadrupole-order term to the gravitational potential felt by planet c, intruding upon the two-planet-resonant picture of inclination evolution developed by \citet{SpaldingBatygin2016}.

\citet{LaiPu2017} show that a distant and massive third body in a 3-planet system can dynamically couple the inner two, arresting mutual inclination growth in the inner system. \revone{The strength of this coupling is set by the parameter}
\begin{align} \label{eq:lai-pu-prameter}
    \varepsilon_{bc} \equiv \frac{\nu_{cd} - \nu_{bd}}{\nu_{bc} + \nu_{cb}} \ll 1,
\end{align}
\revone{which sets the maximum mutual inclination of the inner planets $\theta_{bc}$ as a function of the outer giant's inclination. To leading order in $\theta_{bc}$,
\begin{align}
    \norm{\sin(\theta_{bc,{\rm max}})} \simeq \varepsilon_{bc} \norm{\sin(2I_d)}.
\end{align}
Because both planets b and c are transiting the maximum possible value of $\theta_{bc}$ is a few degrees, and $\sin(\theta_{bc,{\rm max}}) \lesssim 10^{-2}$.}
The precessional frequencies in Equation \eqref{eq:lai-pu-prameter} are
\begin{align}
    \nu_{bc} &= \frac{\bigg m_b m_c a_b}{4a_c^2 \Lambda_b} b_{3/2}^{(1)}\!\left ( \frac{a_b}{a_c} \right ) \\
    \nu_{cb} &= \frac{\bigg m_b m_c a_b}{4a_c^2 \Lambda_c} b_{3/2}^{(1)}\!\left ( \frac{a_b}{a_c} \right ) \\
    \nu_{bd} &= \frac{3\bigg m_b m_d a_b^2}{4a_d^3 \Lambda_b} \\
    \nu_{cd} &= \frac{3\bigg m_b m_d \sqrt{a_b a_c^3}}{4a_d^3 \Lambda_b},
\end{align}
where we have expanded Laplace coefficients pertaining to the outer giant to quadrupole order: $b_{3/2}^{(1)}(\alpha) \simeq 3\alpha$ \citep[e.g.,][]{MurrayDermott1999}. Using the median values we have obtained from \exofast, and an estimated mass of roughly $11\mearth$ for planet c from the mass--radius relation of \citet{ChenKipping2017},
we obtain $\varepsilon_{bc} \sim 10^{-4}$, indicating that this potential third planet strongly couples the orbits of the inner two \revone{regardless, effectively, of its inclination relative to the inner system's orbital plane.} Inclination growth is suppressed by secular torques, therefore, even if the resonant $J_2$ has been achieved.\footnote{\revone{Note that we have adopted here the median minimum mass of the third planet from our \exofast~fit as $m_d$, which is very poorly constrained. Because $\varepsilon_{bc} \propto m_d$, this result depends strongly on $m_d$, but the coupling would be strong for $m_d$ within a few orders of magnitude of $M_J$.}}

We note the analogy to the only other known HJ with a close outer companion: WASP-47. Both planetary systems bear an outer giant of greater mass than the HJ at $P \sim 600$ days. As outlined above, these outer giants may be responsible for coupling the orbits in the inner system and rendering the close companions detectable. If TOI-4468.03 is eventually confirmed as a real planet, the similarity of the TOI-4468 system to the WASP-47 system may allow for additional insights towards its likely formation history, including an assessment of whether the two-stage formation process proposed in \citet{Weiss2017}---in which the giant planets form out of the initial gas-rich protoplanetary disk and are followed by the later \textit{in situ} formation of the sub-Neptune when the disk is gas-poor---could have operated in the TOI-4468 system. 

\section{Conclusions} \label{sec:conclusions}
TOI-4468 is a transiting planetary system comprising a hot Jupiter and a lone, mini-Neptune-sized outer companion. Using light curves from TESS and ground-based observatories, and RV measurements taken with the NEID spectrograph, we have performed a global characterization of this system. Additionally, we have considered the possibility of an additional planet in the system and the consequences of the small mutual inclination between the hot Jupiter and its outer companion. We conclude that:
\begin{itemize}
    \item The TOI-4468 planetary system features a hot Jupiter (planet b) on a 2.77-day orbit, with a close outer companion (planet c) on a 7.01-day orbit and no observed inner companion, the only known system with this architecture.
    \item TOI-4468 b has a mass of 0.54 $M_J$ and radius 1.01 $R_J$, from a joint light curve and radial velocity fit. TOI-4468 c has a radius of 0.28 $R_J$, but we do not constrain its mass.
    \item We identify an additional RV signal which may be due to an outer giant with minimum mass 0.98 $M_J$ on a 624-day orbit. We cannot, however, rule out stellar activity as the source of this signal; more data are needed.
    \item We cannot rule out the existence of a planet interior to planet b. However, any undetected planet \revone{would need to be $\lesssim \rearth$ in size.}
    \item The resonance mechanism invoked to explain the paucity of outer companions to those hot Jupiters which did not undergo high-eccentricity migration may have been thwarted in this system by two possible means. Either the stellar $J_2$ never reached the necessary value to achieve this resonance, or the possible outer giant in the system dynamically coupled their orbits.
\end{itemize}

\begin{acknowledgments}
\revone{The corresponding author is grateful to Rich Townsend and Daniel Jontof-Hutter for useful conversations.} We thank the authors of \citet{Schulte2024, Schulte2025} for making plotting scripts publicly available that informed parts of our visualizations. \revone{We are grateful to the anonymous reviewer, whose suggested revisions improved the clarity of this manuscript and the reproducibility of our analysis.}

J.R.L. is supported by the TESS Guest Investigator program (Grant No. 80NSSC25K7901). This work was funded in part by the University of Wisconsin--Madison Research Forward program sponsored by the Office of the Vice Chancellor for Research (OVCR) through funding provided by the Wisconsin Alumni Research Foundation (WARF). B.J.H. is supported by an appointment to the NASA Postdoctoral Program at NASA Goddard Space Flight Center, administered by Oak Ridge Associated Universities under contract with NASA  (ORAU-80HQTR21CA005).
C.X.H. is supported by ARC Future Fellowship FT240100016.
K.A.C. acknowledges support from the TESS mission via subaward s3449 from MIT.
I.A.S. acknowledges the support of M.V. Lomonosov Moscow State University Program of Development.
The authors acknowledge support from the Swiss NCCR PlanetS and the Swiss National Science Foundation. This work has been carried out within the framework of the NCCR PlanetS supported by the Swiss National Science Foundation under grants 51NF40182901 and 51NF40205606. Funding for K.B. was provided by the European Union (ERC AdG SUBSTELLAR, GA 101054354). J.K. acknowledges support from the Swedish Research Council (Project Grant 2017-04945 and 2022-04043) and of the Swiss National Science Foundation under grant number TMSGI2\_211697.
H.P. acknowledges support by the Spanish Ministry of Science and Innovation with the Ram\'on y Cajal fellowship number RYC2021-031798-I. Funding from the University of La Laguna and the Spanish Ministry of Universities is acknowledged.
A.J. would like to thank the Peter Livingston Scholars Program at the University of Wisconsin--Madison for their generous support. T. M. acknowledges support from the National Science Foundation Graduate Research Fellowship Program (Grant No. DGE-2146755).

Based in part on data obtained by the NEID spectrograph built by Penn State University and operated at the WIYN Observatory by NOIRLab, under the NN-EXPLORE partnership of the National Aeronautics and Space Administration and the National Science Foundation. The NEID archive is operated by the NASA Exoplanet Science Institute at the California Institute of Technology. We thank the NEID Queue Observers and WIYN Observing Associates for their skillful execution of our observations. Based on observations at NSF Kitt Peak National Observatory, NSF NOIRLab, which is managed by the Association of Universities for Research in Astronomy (AURA) under a cooperative agreement with the U.S. National Science Foundation. The authors are honored to be permitted to conduct astronomical research on I’oligam Du’ag (Kitt Peak), a mountain with particular significance to the Tohono O’odham.

This work makes use of observations from the LCOGT network. Part of the LCOGT telescope time was granted by NOIRLab through the Mid-Scale Innovations Program (MSIP). MSIP is funded by NSF.

This research has made use of the Exoplanet Follow-up Observation Program (ExoFOP; DOI: 10.26134/ExoFOP5) website, which is operated by the California Institute of Technology, under contract with the National Aeronautics and Space Administration under the Exoplanet Exploration Program.

Funding for the TESS mission is provided by NASA's Science Mission Directorate.
\end{acknowledgments}

\section*{Data Availability}
The light curve data presented in this article were obtained from the Mikulski Archive for Space Telescopes (MAST) at the Space Telescope Science Institute. The specific observations analyzed can be accessed via \dataset[doi:10.17909/6fdy-r386]{https://doi.org/10.17909/6fdy-r386}. The detrended TESS light curves are available on Zenodo: \href{https://doi.org/10.5281/zenodo.20029593}{10.5281/zenodo.20029593}. Speckle images and light curves from ground-based observatories are available on ExoFOP: \href{https://doi.org/10.26134/ExoFOP5}{10.26134/ExoFOP5}. The processed RV data we collected with NEID are reported in Table \ref{tab:rvs}.

\facilities{TESS, WIYN, LCOGT, RFAC, SUTO}

\software{\exofast~\citep{Eastman2019}, \triceratops~\citep{triceratops-paper, triceratops-code}, \vplanet~\citep{Barnes2020}, \batman~\citep{Kreidberg2015}, \codestyle{numpy}~\citep{Harris2020}, \codestyle{matplotlib}~\citep{Hunter2007}, \codestyle{astropy}~\citep{astropy2022}, \codestyle{pandas} \citep{pandas-paper}, \astroimagej~\citep{Collins:2017}, \codestyle{tapir}~\citep{Jensen:2013}}

\providecommand{\bjdtdb}{\ensuremath{\rm {BJD_{TDB}}}}
\providecommand{\tjdtdb}{\ensuremath{\rm {TJD_{TDB}}}}
\providecommand{\feh}{\ensuremath{\left[{\rm Fe}/{\rm H}\right]}}
\providecommand{\teff}{\ensuremath{T_{\rm eff}}}
\providecommand{\teq}{\ensuremath{T_{\rm eq}}}
\providecommand{\ecosw}{\ensuremath{e\cos{\omega_*}}}
\providecommand{\esinw}{\ensuremath{e\sin{\omega_*}}}
\providecommand{\lsun}{\ensuremath{\,L_\Sun}}
\providecommand{\mj}{\ensuremath{\,M_{J}}}
\providecommand{\rj}{\ensuremath{\,R_{J}}}
\providecommand{\fave}{\langle F \rangle}
\providecommand{\fluxcgs}{10$^9$ erg s$^{-1}$ cm$^{-2}$}

\startlongtable

\begin{deluxetable*}{lccccccc}
\tabletypesize{\footnotesize}
\tablecaption{Stellar and planetary properties and RV instrument characteristics. Median values and 68\% confidence intervals from \exofast.}
\tablehead{\colhead{~~~Parameter} & \colhead{Description} & \multicolumn{3}{c}{Values}}
\startdata
\smallskip\\\multicolumn{2}{l}{Fitted Stellar Parameters:}&\smallskip\\
~~~~$T_{\rm eff}$\dotfill &Effective temperature (K)\dotfill &$5170^{+120}_{-110}$\\
~~~~$\log_{10}(g)$\dotfill &Surface gravity (cgs)\dotfill &$4.571^{+0.020}_{-0.022}$\\
~~~~$F_{\rm bol}$\dotfill &Bolometric flux (cgs)\dotfill &$8.53^{+0.82}_{-0.64} \times 10^{-11}$\\
~~~~$[{\rm Fe/H}]$\dotfill &Metallicity (dex)\dotfill &$-0.03\pm0.16$\\
~~~~$\varpi$\dotfill &Parallax (mas)\dotfill &$2.645\pm0.013$\\
~~~~$d$\dotfill &Distance (pc)\dotfill &$378.1\pm1.9$\\
\smallskip\\\multicolumn{2}{l}{Derived Stellar Parameters:}&\smallskip\\
~~~~$M_\star$\dotfill &Mass ($\msun$)\dotfill &$0.805^{+0.042}_{-0.038}$\\
~~~~$R_\star$\dotfill &Radius ($\rsun$)\dotfill &$0.771^{+0.020}_{-0.018}$\\
~~~~$L_\star$\dotfill &Bolometric luminosity ($\lsun$)\dotfill &$0.381^{+0.036}_{-0.029}$\\
~~~~$\rho_\star$\dotfill &Density (g cm$^{-3}$)\dotfill &$2.49^{+0.15}_{-0.17}$\\
~~~~Age\dotfill &Age (Gyr)\dotfill &$7.4^{+3.9}_{-3.6}$\\
~~~~$A_V$\dotfill &V-band extinction (mag)\dotfill &$0.22^{+0.14}_{-0.13}$\\
\smallskip\\\multicolumn{2}{l}{Fitted Planetary Parameters:}&TOI-4468 b&TOI-4468 c&TOI-4468.03\tablenotemark{a}\smallskip\\
~~~~$P$\dotfill &Period (days)\dotfill &$2.77085931\pm0.00000036$&$7.013900\pm0.000020$&$624^{+53}_{-38}$\\
~~~~$K$\dotfill &RV semi-amplitude (m/s)\dotfill &$89.4^{+3.9}_{-3.5}$&---&$25.8^{+4.5}_{-46}$\\ 
~~~~$R_p/R_\star$\dotfill &Radius of planet in stellar radii \dotfill &$0.13424^{+0.00062}_{-0.00054}$&$0.0367^{+0.0013}_{-0.0011}$&---\\
~~~~$a/R_\star$\dotfill &Semi-major axis in stellar radii \dotfill &$10.03^{+0.20}_{-0.23}$&$18.63^{+0.37}_{-0.43}$&$371^{+22}_{-17}$\\
~~~~$e\cos(\omega)$ & \dotfill &$0.015^{+0.024}_{-0.016}$&---&---\\
~~~~$e\sin(\omega)$ & \dotfill &$0.016^{+0.024}_{-0.017}$&---&---\\
~~~~$m_p\sin(i)$ &Minimum mass ($\mearth$)\dotfill &$169.6^{+9.8}_{-8.8}$&---&$311^{+49}_{-51}$\\ 
~~~~$i$\dotfill &Inclination (deg)\dotfill &$89.35^{+0.44}_{-0.49}$&$88.95^{+0.66}_{-0.50}$&---\\
~~~~$T_C$\dotfill &Observed time of conjunction\dotfill
&$2848.554232^{+0.000077}_{-0.000076}$&$2424.2354^{+0.0025}_{-0.0022}$&$3736^{+220}_{-23}$\\
~~~~$T_{14}$\dotfill &Total transit duration (days)\dotfill &$0.09743^{+0.00036}_{-0.00033}$&$0.1156^{+0.0030}_{-0.0032}$\\
\smallskip\\\multicolumn{2}{l}{Derived Planetary Parameters:}&TOI-4468 b&TOI-4468 c&TOI-4468.03\smallskip\\
~~~~$R_p$\dotfill &Radius ($\rearth$)\dotfill &$11.29^{+0.31}_{-0.28}$&$3.09^{+0.14}_{-0.12}$&---\\
~~~~$m_p$\dotfill &Mass ($\mearth$)\dotfill &$170.1^{+9.7}_{-8.8}$&$11.1^{+4.0}_{-2.6}$~\tablenotemark{b}&---\\ 
~~~~$a$\dotfill &Semi-major axis (au)\dotfill &$0.03592^{+0.00060}_{-0.00057}$&$0.0667\pm0.0011$&$1.332^{+0.079}_{-0.061}$\\
~~~~$e$\dotfill &Eccentricity \dotfill &$0.030^{+0.026}_{-0.020}$&$0.122^{+0.16}_{-0.083}$&---\\
~~~~$\omega$\dotfill &Argument of periastron (deg)\dotfill &$46^{+45}_{-47}$&$170^{+120}_{-140}$&---\\
~~~~$T_{\rm eq}$\dotfill &Equilibrium temp (K)\dotfill &$1153^{+26}_{-22}$&$846^{+19}_{-16}$&$189.5^{+6.3}_{-6.4}$\\
~~~~$b$\dotfill &Transit impact parameter \dotfill &$0.111^{+0.084}_{-0.076}$&$0.33^{+0.20}_{-0.21}$&---\\
~~~~$\rho_p$\dotfill &Density (cgs)\dotfill &$0.651^{+0.055}_{-0.053}$&$2.06^{+0.74}_{-0.47}$&---\\
~~~~$\log_{10}(g_p)$\dotfill &Surface gravity (cgs)\dotfill &$3.117\pm0.028$&$3.05^{+0.13}_{-0.11}$&---\\
~~~~$\fave$\dotfill &Incident flux (\fluxcgs)\dotfill &$0.401^{+0.037}_{-0.030}$&$0.1131^{+0.011}_{-0.0095}$&$0.000292^{+0.000041}_{-0.000037}$\\
\smallskip\\\multicolumn{2}{l}{Telescope Parameters:}&NEID\smallskip\\
~~~~$\gamma_{\rm rel}$\dotfill &Relative RV offset (m/s)\dotfill &$-19.9^{+2.4}_{-3.0}$\\
~~~~$\sigma_J$\dotfill &RV jitter (m/s)\dotfill &$0.0^{+6.2}_{-0.00}$\\
~~~~$\sigma_J^2$\dotfill &RV jitter variance \dotfill &$-12^{+51}_{-17}$\\
\enddata

\label{tab:exofast}

\tablenotetext{a}{We cannot confirm the planetary nature of this radial velocity signal. We include planetary parameters assuming this scenario for completeness.}
\tablenotetext{b}{We report a mass, density, and surface gravity for planet c based on the mass--radius relation of \citet{ChenKipping2017}, though we cannot establish a mass measurement from our measured radial velocities.}

\end{deluxetable*}

\providecommand{\bjdtdb}{\ensuremath{\rm {BJD_{TDB}}}}
\providecommand{\tjdtdb}{\ensuremath{\rm {TJD_{TDB}}}}
\providecommand{\feh}{\ensuremath{\left[{\rm Fe}/{\rm H}\right]}}
\providecommand{\teff}{\ensuremath{T_{\rm eff}}}
\providecommand{\teq}{\ensuremath{T_{\rm eq}}}
\providecommand{\ecosw}{\ensuremath{e\cos{\omega_*}}}
\providecommand{\esinw}{\ensuremath{e\sin{\omega_*}}}
\providecommand{\lsun}{\ensuremath{\,L_\Sun}}
\providecommand{\mj}{\ensuremath{\,M_{\rm J}}}
\providecommand{\rj}{\ensuremath{\,R_{\rm J}}}
\providecommand{\me}{\ensuremath{\,M_{\rm E}}}
\providecommand{\re}{\ensuremath{\,R_{\rm E}}}
\providecommand{\fave}{\langle F \rangle}
\providecommand{\fluxcgs}{10$^9$ erg s$^{-1}$ cm$^{-2}$}

\startlongtable

\begin{deluxetable*}{lccc}
\tabletypesize{\footnotesize}
\tablecaption{Stellar and instrumental photometric characteristics. Median values and 68\% confidence intervals from \exofast.}
\tablehead{\colhead{~~~Parameter} & \colhead{Description} & \multicolumn{2}{c}{Values}}
\startdata
\smallskip\\\multicolumn{2}{l}{Wavelength Parameters:}&$B$&$R$\smallskip\\
~~~~$u_{1}$\dotfill &Linear limb-darkening coeff \dotfill &$0.811^{+0.063}_{-0.066}$&$0.473^{+0.046}_{-0.047}$\\
~~~~$u_{2}$\dotfill &Quadratic limb-darkening coeff \dotfill &$0.024\pm0.063$&$0.202\pm0.041$\\
\smallskip\\&&$i'$&TESS\smallskip\\
~~~~$u_{1}$\dotfill &Linear limb-darkening coeff \dotfill &$0.423\pm0.056$&$0.408^{+0.024}_{-0.025}$\\
~~~~$u_{2}$\dotfill &Quadratic limb-darkening coeff \dotfill &$0.234^{+0.052}_{-0.051}$&$0.214\pm0.037$\\
\smallskip\\\multicolumn{2}{l}{Transit Parameters:}&TESS Full Frame Images (TESS)&TESS 2-Minute Cadence (TESS)\smallskip\\
~~~~$\sigma^{2}$\dotfill &Added Variance \dotfill &$3.5^{+1.1}_{-1.0} \times 10^{-7}$&$4.86\pm0.27 \times 10^{-6}$\\
~~~~$F_0$\dotfill &Baseline flux \dotfill &$0.999952\pm0.000037$&$1.000080\pm0.000029$\\
\smallskip\\&&SUTO UT 2022-04-09 ($B$)&RFAC UT 2023-01-11 ($R$)\smallskip\\
~~~~$\sigma^{2}$\dotfill &Added Variance \dotfill &$-2.02^{+0.98}_{-0.72} \times 10^{-5}$&$-3.22^{+0.70}_{-0.52} \times 10^{-5}$\\
~~~~$F_0$\dotfill &Baseline flux \dotfill &$0.99904^{+0.0010}_{-0.00100}$&$0.99890^{+0.00082}_{-0.00081}$\\
\smallskip\\&&RFAC UT 2023-04-07 ($R$)&LCO UT 2024-08-16 ($i'$)\smallskip\\
~~~~$\sigma^{2}$\dotfill &Added Variance \dotfill &$-0.0003131^{+0.000011}_{-0.0000090}$&$0.5^{+2.9}_{-2.5} \times 10^{-7}$\\
~~~~$F_0$\dotfill &Baseline flux \dotfill &$0.99978^{+0.00092}_{-0.00090}$&$1.00010\pm0.00014$\\
\enddata

\label{tab:exofast-photometry}

\end{deluxetable*}

\bibliography{refs}{}
\bibliographystyle{aasjournal}


\end{document}